\documentclass[%
 reprint,
superscriptaddress,
 amsmath,amssymb,
 aps,
]{revtex4-2}
\usepackage{xcolor}
\usepackage{physics}
\usepackage{graphicx}
\usepackage{dcolumn}
\usepackage{bm}


\newcommand{\1}{{\rm 1\hspace{-0.9mm}l}}

\usepackage[colorlinks=true,linkcolor=blue,urlcolor=blue,citecolor=blue,pdfusetitle]{hyperref}
\usepackage{hyperref}

\begin{document}

\preprint{APS/123-QED}

\title{Salient signatures of entanglement in the surrounding environment}

\date{\today}

\author{{\L}ukasz Rudnicki}
\affiliation{International Centre for Theory of Quantum Technologies, University of Gda{\'n}sk, 80-308 Gda{\'n}sk, Poland}

\author{Waldemar K{\l}obus}
\affiliation{Institute of Theoretical Physics and Astrophysics, Faculty of Mathematics, Physics, and Informatics, University of Gda{\'n}sk, 80-308 Gda{\'n}sk, Poland}

\author{Otavio A. D. Molitor}
\affiliation{International Centre for Theory of Quantum Technologies, University of Gda{\'n}sk, 80-308 Gda{\'n}sk, Poland}

\author{Wies{\l}aw Laskowski}
\affiliation{International Centre for Theory of Quantum Technologies, University of Gda{\'n}sk, 80-308 Gda{\'n}sk, Poland}
\affiliation{Institute of Theoretical Physics and Astrophysics, Faculty of Mathematics, Physics, and Informatics, University of Gda{\'n}sk, 80-308 Gda{\'n}sk, Poland}

\begin{abstract}
We develop a model in which presence of entanglement in a quantum system can be confirmed through coarse observations of the environment surrounding the system. This counter-intuitive effect becomes possible when interaction between the system and its environment is proportional to an observable being an entanglement witness. While presenting \textcolor{black}{intuitive examples we show that: i) a cloud of an ideal gas, when subject to a linear potential coupled with the entanglement witness, accelerates in the direction dictated by the sign of the witness; ii) when the environment is a radiation field, the direction of dielectric polarization depends on the presence of entanglement; iii) quadratures of electromagnetic field in a cavity coupled with two qubits (or a four-level atom) are displaced in the same manner.}
\end{abstract}

\maketitle


\section{Introduction}

Entanglement~\cite{Einstein1935,Bell1964,Horodecki2009} is considered to be a very versatile feature of quantum systems~\cite{Ekert1991,Nielsen2000,Jozsa2003,Renner2005}. Therefore, while interaction of an entangled system with yet another quantum system can reveal information about entanglement in the former system, it is doubtful that the same might happen while interacting with an environment~\cite{Rajagopal2001,Zyczkowski2001}. To account for huge discrepancies between both, visible in the size (number of degrees of freedom) and typical complexity of the environment, one usually assumes a very realistic and physically motivated Markovian approximation, in which the environment is memoryless (its correlations decay very fast with time and no information flows back to the system)~\cite{Breuer2002}. In practice, while one always encounters some residual non-Markovian (memory) effects~\cite{Bellomo2007,Breuer2016}, we expect that information about the quantum system spreads over the environment in an uncontrolled way, eventually imprinting very cumbersome and extremely weak remains~\cite{Zurek2003,Schlosshauer2005,Davidovich2007}. We may fairly expect that even full control over all degrees of freedom of the environment (even though under this circumstance it does not anymore make sense to speak about environment) would not generally give us means to read out information about entanglement of the quantum system interacting with it~\cite{Romano2006}.

In this letter we challenge that quite natural perspective, proposing a scheme in which entanglement in the system can very clearly affect the evolution of coarse degrees of freedom of the environment. The gist lays in a notion of an entanglement witness --- an observable which is positive for all separable states~\cite{Kai2004,Guhne2009,Chruscinski2014,Chruscinski2018}. While it seems hopeless to look for entanglement scrutinizing  details of environment's evolution, we introduce a model of system-environment interaction in which the \textit{direction} of the evolution depends on entanglement in the system.




To be more precise, let us assume that a quantum system in question is \textcolor{black}{coupled to a second system, which we call the environment of the former \cite{Breuer2002}. This universal definition covers both ``proper'' (thermal) reservoirs, for which all the limitations listed above occur, and ancillary quantum systems as simple as a single qubit (useful e.g. as a probe in metrology \cite{Chu2021}). However, while we develop the theory framework in general, in the examples' part we only consider the reservoirs.} Can we certify entanglement inside the system just by investigating some features of the environment? We answer this question affirmatively, so that in principle we can confirm entanglement in the system without the need to perform measurements directly, in which case the state of the system (and entanglement treated as a valuable resource) remains intact. 
With this respect, our approach intends to utilize an additional medium through which we aim to extract useful information, in contrast to  performing non-demolition \cite{wm1,wm2,wm3} or protective quantum measurements \cite{pm1,pm1}.

On the other hand, it has been shown that the additional medium avails generation of entanglement in distinct quantum system \cite{ancm1,ancm2}, possibly also through reservoir engineering in 
open quantum systems \cite{res1,res2,res3}.
However, we shall stress that we are neither concerned with entanglement between the system and the environment, which occurs naturally due to interaction between the two, nor in macroscopic entanglement, present e.g. in large spin systems \cite{Wiesniak_2005}, where the environment plays no role. We rather wish to make a few observations of the sole environment, and be able to decide about presence of entanglement only in the system interacting with it.

In the following we introduce the physical model given by the total Hamiltonian $H$ describing the system $S$ (the entanglement of which is to be investigated) interacting with its environment $E$. The interaction Hamiltonian $H_{\textrm{int}}$, present in $H$, is selected in such a way that: i) it is proportional to an entanglement witness acting on $S$; ii) it highly influences time evolution of certain environment observables. The latter are at the end measured (or simply observed) in order to infer possible existence of entanglement within $S$. In regard to this step, we note that entanglement in the system is not to be certified by a particular outcome of a measurement relevant for environment observables with discrete spectra, but rather, by coarse features thereof. Moreover, we assure constancy of the entanglement witness (in the Heisenberg picture), which means that if the system is initially entangled, the discussed time evolution will not destroy it completely.

In the next step we present the details of the \textcolor{black}{protocol. Then, after setting the scene, we discuss: i)  a toy model of an ideal gas; ii) quantized radiation field as a benchmark for a reservoir; iii) single-mode field interacting with a four-level atom (akin of two entangled qubits). 
}

\begin{figure}
	\includegraphics[width=\columnwidth]{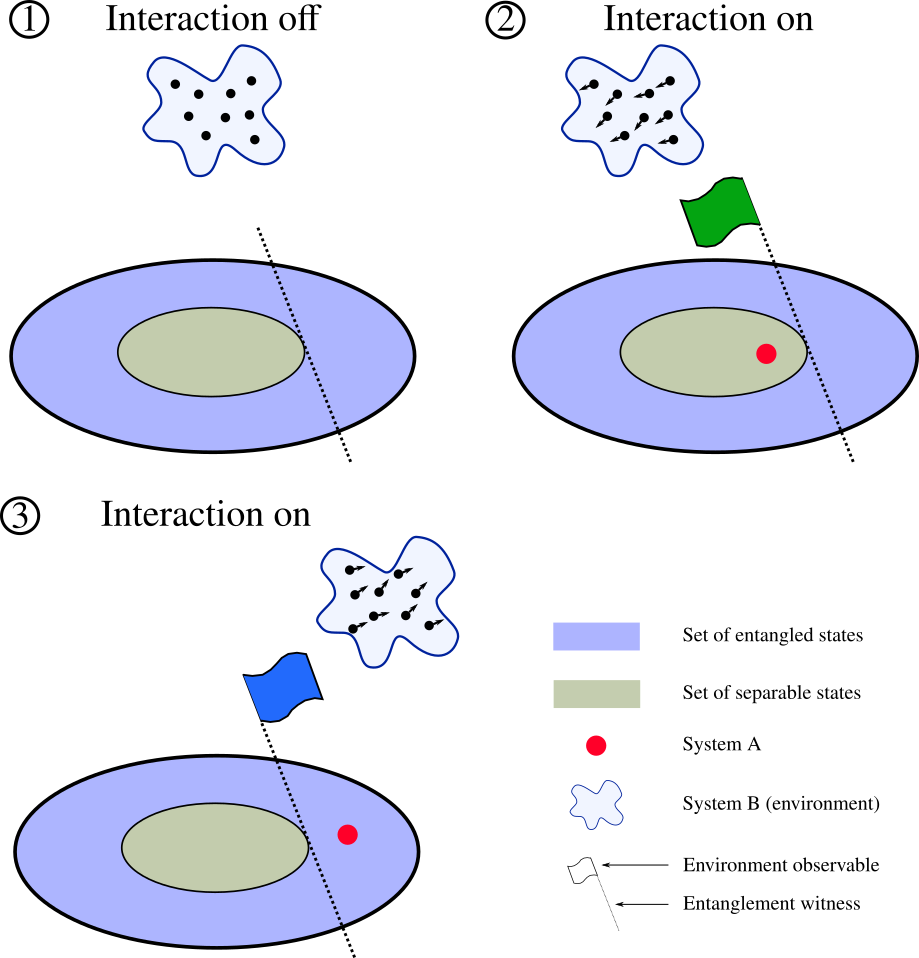}
\caption{Entanglement witness (represented as a dashed line) couples the system (shown as a red dot) with its surrounding environment (here depicted as a gas of particles). \raisebox{.5pt}{\textcircled{\raisebox{-.9pt} {1}}} --- The interaction is switched off. The environment does not reveal the nature of the system, it might be in the entangled set of states, or not (blue and beige regions, respectively). \raisebox{.5pt}{\textcircled{\raisebox{-.9pt} {2}}} and \raisebox{.5pt}{\textcircled{\raisebox{-.9pt} {3}}} --- The interaction is switched on. The gas moves following the flag, while the flag shows the value of the entanglement witness: green flag to the left corresponds to a separable state; blue flag to the right indicates entanglement.} \label{obr}
\end{figure}

\section{The protocol} 

Let the state $\rho\left(t\right)$ represent a composite physical system which consists of system $S$ coupled with
the environment $E$, together treated as a closed system governed
by the evolution equation (we set $\hbar=1$)
\begin{equation}
\frac{d}{dt}\rho\left(t\right)=i\left[\rho\left(t\right),H\right].
\end{equation}
The Hamiltonian of the total system $H$ is
\begin{equation}
H=H_{S}\otimes\1_{E}+\1_{S}\otimes H_{E}+H_{\textrm{int}},
\end{equation}
where $H_S$ and $H_E$ are  acting on Hilbert spaces of $S$ and $E$ respectively, whereas $H_{\textrm{int}}$ describes the interaction of the system $S$ with the environment $E$. The system $S$ is by itself assumed to be a composite one (e.g. two qubits), so that its state $\rho_S (t)=\mathrm{tr}_E \rho(t)$ might be entangled. 

Moreover, let $\{G_{j}\}_{j=1}^K$ be a set consisting of an arbitrary number of $K$ time-independent observables of the environment, and denote
\begin{equation}
G_{j}\left(t\right)=\tr\left\{(\1_{S}\otimes G_{j})\rho\left(t\right)\right\}.
\end{equation}
These are time-dependent average values of the environment observables. Our precise goal is to explore the presence of entanglement within $\rho_S (t)$ by probing the qualitative behavior of the observables $G_{j}\left(t\right)$, as presented schematically in Fig. \ref{obr}.


In order to make it possible, we obviously need interaction
between the system and the environment which would mediate entanglement of the system to the environment. To this end we set the interaction to be
\begin{equation}\label{haint}
H_{\textrm{int}}=W\otimes h_{E},
\end{equation}
where $W$ is arbitrarily chosen \emph{entanglement witness} for the system $S$, while
$h_{E}$ is an operator of the environment to be specified later on. Thus we have
\begin{equation}
W\left(t\right)=\tr\left\{(W\otimes\1_{E})\rho\left(t\right)\right\},
\end{equation}
which, whenever assumes a negative value, proves entanglement of $\rho_S (t)$. In this way we let the coupling between the system $S$ with its environment $E$ depend on the presence of entanglement in $S$. To our best knowledge such a proposal has not yet been considered.

Note that for the entanglement witness operator $W\otimes\1_{E}$, defined on the whole $S+E$ system, we have  
\begin{equation}
\left[W\otimes\1_{E},H\right]=\left[W,H_{S}\right]\otimes\1_{E}.
\end{equation}
In other words, only the system Hamiltonian $H_{S}$ in the first infinitesimal time step governs the evolution of  $W(t)$.
Therefore, if  $\left[W,H_{S}\right]=0$, then $W\left(t\right)$ is \textcolor{black}{``frozen''}.
In such a model, even though the state of the system evolves, its entanglement does not vanish in a sense that if detected once, it will always lead to negative $W(t)$. In what follows we assume that $\left[W,H_{S}\right]=0$, and consequently that $W(t)\equiv W_0$ is time-independent.

Using the evolution equation we can derive equations of motion for the observables $G_j$
\begin{subequations}
\label{rownaG}
\begin{equation}
    \dot{G}_{j}(t)=i\,\tr\left\{\left[H,\1_{S}\otimes G_{j}\right]\rho\left(t\right)\right\} =F_E^{(j)}+F_W^{(j)},
\end{equation}
where environment-and-witness-origin ``forces'' are:
\begin{equation}
    F_E^{(j)}=i\,\tr\left\{\1_{S}\otimes\left[H_{E},G_{j}\right]\rho\left(t\right)\right\},
\end{equation}
\begin{equation}
    F_W^{(j)}=i\,\tr\left\{W\otimes\left[h_{E},G_{j}\right]\rho\left(t\right)\right\}.
\end{equation}
\end{subequations}
While these are not forces as to be understood in classical mechanics (with an exception of $G_j$ being kinetic momenta), we adopt this name here and from now on we abandon quotation marks.

Every $F_E^{(j)}$ depends solely on the state of the environment $\rho_E (t)=\mathrm{tr}_S \rho(t)$, which means that if it contains information about entanglement in the system $S$, this information is hidden inside correlations built up in the environment along the time evolution. This corresponds to a generic scenario already outlined in the Introduction. Therefore, we do not expect it to be a useful resource of information and we rather wish to mitigate the supposedly counterproductive influence of these forces. To this end we assume that the commutator $\left[H_{E},G_{j}\right]$ depends solely on the chosen observables $\left\{ G_{j}\left(t\right)\right\} $, i.e. 
\begin{equation}
\left[H_{E},G_{j}\right]=i\sum_{k}c_{jk}G_{k}.
\end{equation}
Note that the constants $c_{jk}$ are real because the commutator is anti-hermitian, while $H_{E}=H_{E}^\dagger$ and $G_{j}=G_{j}^\dagger$.

In order to extract potentially useful and accessible information about the entanglement in $S$, we resort to the witness type of forces. If we assume that at the initial time $t=0$ both $S$ and $E$ are mutually uncorrelated, $\rho\left(0\right)=\rho_{S}\left(0\right)\otimes\rho_{E}\left(0\right)$, the total system evolves as
\begin{equation}
    \rho\left(t\right)=e^{-iHt}\rho_{S}\left(0\right)\otimes\rho_{E}\left(0\right)e^{iHt}.
\end{equation}

Now, we can express the entanglement witness operator in its eigenbasis $W=\sum_{k}w_{k}\left|\chi_{k}\right\rangle \left\langle \chi_{k}\right|$, so that the expectation value reads
\begin{equation}
    W_{0}=\tr\left\{ (W\otimes \1_E) \rho(t) \right\}=\sum_{k}w_{k}\left\langle \chi_{k}\right|\rho_{S}\left(0\right)\left|\chi_{k}\right\rangle,
\end{equation}
because, by assumption, $W$ commutes with the total Hamiltonian.
This also allows us to write
\begin{eqnarray}\label{Wforces}
\lefteqn{\tr\left\{W\otimes\left[h_{E},G_{j}\right]\rho\left(t\right)\right\}=\sum_{k}w_{k}\left\langle \chi_{k}\right|\rho_{S}\left(0\right)\left|\chi_{k}\right\rangle}  \\
&& \times\tr\left\{e^{it\left(H_{E}+w_{k}h_{E}\right)}\left[h_{E},G_{j}\right]e^{-it\left(H_{E}+w_{k}h_{E}\right)}\rho_{E}\left(0\right)\right\}.\nonumber
\end{eqnarray}

At this point one can see that further investigation depends on the particular selection of measurement operators in relation to the environment part of interaction Hamiltonian, so that the influence of entanglement witness in the evolution of $G_{j}\left(t\right)$ becomes straightforward to analyse.
Evidently, the easiest option would be to let
\begin{equation}\label{gj}
\left[h_{E},G_{j}\right]=ig_{j}\1_{E},
\end{equation}
with all constants $g_{j}$ again being real. However, as this requirement cannot be fulfilled by operators with discrete and bounded spectra, it shall be adopted with caution.

For continuous and unbounded observables of the environment, with the help of (\ref{gj}), the witness force reduces to a constant $ig_{j}$, so that the equations of motion  read
\begin{equation}\label{eqnss}
\dot{G}_{j}\left(t\right)+\sum_{k}c_{jk}G_{k}\left(t\right)+g_{j}W_{0}=0.
\end{equation}
As one with constant coefficients, the system \eqref{eqnss} can formally be solved in a general case. However, since finding a solution does not imply knowledge about the sign of $W_0$, we further look for scenarios in which, as promised in the title, the solutions do depend on this sign in a salient way. Moreover, we also note in passing that Eq. (\ref{eqnss}) can be reverted in order to give the entanglement witness as a function of the environment variables. Therefore, being able to track all the observables $\left\{ G_{j}\left(t\right)\right\}$,
we can infer the value of $W_{0}$. 

In the following, we discuss a few selected scenarios which allow to provide the sign of the  entanglement witness in terms of salient qualitative effects.

\section{Environment as a cloud of ideal gas} 

First, let us illustrate our method by looking at the environment being  a gas of identical particles with mass $m$.
Since a generalization to more spatial dimensions is straightforward, for clarity let us consider 1-dimensional ideal gas of $N$ particles
with coordinates $\left(q_{1},p_{1},\ldots,q_{N},p_{N}\right)$, so that the Hamiltonian of the environment takes the form
\begin{equation}
H_{E}=\frac{1}{2m}\sum_{k=1}^{N}p_{k}^{2}.
\end{equation}
We set $K=1$ and select our single observable to be the center-of-mass momentum
\begin{equation}
G_{1}=\frac{1}{N}\sum_{k=1}^{N}p_{k} =: P.
\end{equation}
Clearly $\left[P,H_{E}\right]=0$, as the momentum is conserved. We let the interaction depend linearly on the position of every particle (each particle of the environment is coupled to the system in the
same way) $h_{E}=\alpha\sum_{k=1}^{N}q_{k} =: \alpha NR$,
where $R$ is center-of-mass position of the cloud. Such a choice resembles \textcolor{black}{dipole coupling with external electric field, though, here it is rather used as a toy model (interaction} of that kind would need to be engineered in a more elaborate way). Since $\left[h_{E},P\right]=i\alpha \1$, \textcolor{black}{ for any frozen witness $W$}
the equation of motion \eqref{eqnss} becomes
\begin{equation}\label{ideal}
\dot{P}\left(t\right)+\alpha W_{0}=0.
\end{equation}
\textcolor{black}{Assuming $\rho_E(0)$ to be a thermal state $\rho_{E}(0)\sim e^{-\beta H_{E}}$, $P(t)$ coherently with every individual $p_k(t)$} evolves as
\begin{equation}
P\left(t\right)=-\alpha W_{0}t.
\end{equation}
\textcolor{black}{
Neglecting quantum effects in the environment's statistics, the standard deviation of $P(t)$ is $\Sigma_P (t)= \sqrt{\alpha^2 \Sigma_W^2 t^2 +m k_B T/N}$ (check Supplemental Material), where $\Sigma_W$ is the standard deviation of $W$ on $\rho_S(0)$. This confirms that thermal fluctuations severely affecting an individual constituent of the environment ($N=1$), for the center-of-mass motion are damped in the limit of high $N$. Therefore, Eq. (\ref{ideal}) is physically meaningful whenever $\Sigma_W\ll \left|W_0\right|$.
}

The interpretation of this formula is straightforward and at the same time beautiful: the system $S$ is detected to be entangled (i.e. the witness $W_0$ assumes negative values), whenever the center of mass of the  surrounding cloud of gas accelerates towards rising values of position coordinates. If no entanglement is detected, the center of mass of the cloud of gas moves to the opposite side.

\section{A system coupled to a radiation field} 

Let us now consider the environment to be {\color{black} a bath consisting of several electromagnetic modes. In such a case, the Hamiltonian of the environment is given by ($\hbar=1=c$)
\begin{equation}
    H_E = \sum_{\boldsymbol k,\lambda} \omega_{k} \left(a^\dagger_\lambda(\boldsymbol k) a_\lambda(\boldsymbol k)+\frac{1}{2}\right),
\end{equation}
where $\boldsymbol k$ is the wave vector and $\omega_{k} = |\boldsymbol k| = k$  is the frequency corresponding to the mode $\boldsymbol k$. As usual, $a_\lambda(\boldsymbol k)$ and $a^\dagger_\lambda(\boldsymbol k)$ are the annihilation and creation operators of the mode $\boldsymbol k$ with polarization $\lambda$. The above Hamiltonian stems from quantization of the electromagnetic field in terms of operators $\boldsymbol E(\boldsymbol{r})$ and $\boldsymbol B(\boldsymbol{r})$.

We choose the interaction term $h_E$ to be equal to
\begin{equation}\label{dipole}
h_E = -\int \dd^3\boldsymbol{r}\boldsymbol D (\boldsymbol{r})\cdot \boldsymbol E(\boldsymbol{r}),
\end{equation}
with $\boldsymbol D(\boldsymbol{r})$ being a real, transversal (i.e. $\boldsymbol\nabla\cdot \boldsymbol D(\boldsymbol{r})=0$) vector representing the dipole moment. The interaction $H_\mathrm{int}$ consists of a typical dipole term (\ref{dipole}) coupled with $W$. This means that the atomic dipole operator of the system is equal to $W D(\boldsymbol{r})$. While the vector $D(\boldsymbol{r})$ describes its spatial dependence, $W$ is responsible for "internal" degrees of freedom of entangled atoms (e.g. spins, in which case the electric dipole spin resonance might be a method to generate $H_\mathrm{int}$). A simple instance of that scenario is discussed in the next paragraph.


Solving the equations of motion with initial conditions fixed by the thermal state (like in the ideal gas example, but now in the multimode Fock space) we can find the averages of the environment observables (check Supplemental Material): $\boldsymbol E(\boldsymbol{r},t)$ and $\boldsymbol B(\boldsymbol{r},t)$, playing the roles of the macroscopic variables $G_j(t)$. As the final step of coarse graining we average the EM field over time:
\begin{equation}
  \overline{\boldsymbol E(\boldsymbol{r})} =\lim_{T\rightarrow\infty} \frac{1}{T} \int_0^T \dd t \, \boldsymbol E(\boldsymbol{r},t), \qquad \mathrm{etc.},
\end{equation}
getting: $\overline{\boldsymbol B(\boldsymbol{r})} = 0$, and
\begin{equation}
    \overline{\boldsymbol E(\boldsymbol{r})} =  \frac{W_0}{\epsilon_0} \boldsymbol D(\boldsymbol{r}).
\end{equation}
This final result shows that the sign of the entanglement witness dictates the direction of the resulting mean electric field, or rather, spatially localized electric susceptibility. Moreover, if we consider a situation in which instead of a single pair of atoms we deal with a medium composed of many identical entangled pairs (all entangled in their internal degrees of freedom), under suitable experimental conditions, the refractive index of this medium would be sensitive to the presence of entanglement.

We shall emphasize that, as explained in the introduction, it is impossible to keep track of all the degrees of freedom of the radiation environment. Therefore, we just rely on the effective field $ \overline{\boldsymbol E(\boldsymbol{r})}$. 

\section{A four-level atom coupled to a single-mode} 

Although having a clear theoretical description, in the above example the particular form of interaction between the system $S$ and its environment might not always exhibit an intuitive experimental implementation. A potential testbed can be sought in terms of  a four-level system coupled to a single-mode photon field in a cavity~\cite{Albert2012,Xie2017,Zhang2021}. The Hamiltonian governing such a system is \cite{4lev}   }
\begin{equation}\label{Ham4}
      H =\sum_{i=1}^4 \varepsilon_i \sigma_{ii} + \omega a^\dagger a + \gamma(\sigma_{14}+\sigma_{41})(a+ a^\dagger),
\end{equation}
where $a$ and $a^\dagger$ are annihilation and creation operators corresponding to the photon field with frequency $\omega$. Clearly, the observables $G_1$ and $G_2$ become proportional to the quadratures of the field.

Moreover, $\varepsilon_i$ are the four energy levels of the atom, $\gamma$ is the atom–field coupling parameter, while $\sigma_{ij}=\left|i\right\rangle \left\langle j\right|$ for $i,j=1,\ldots,4$ are raising/lowering operators between the levels $|i\rangle$ and $|j\rangle$.

\textcolor{black}{To certify entanglement present in the system of two qubits we can resort to the observable
\begin{equation}\label{wu}
     \tilde W=\sigma_x \otimes \sigma_x - \sigma_y \otimes \sigma_y.
\end{equation}
Entanglement is certified whenever $|\langle \tilde W \rangle|>1$, so we can construct two independent witnesses $W_\pm=\1_4\pm\tilde W$, each sensitive to a neighborhood of a distinct Bell state.}

We further map a composite system of two qubits into a single four-level system. Four basis states: $|00\rangle$, $|01\rangle$, $|10\rangle$, and $|11\rangle$ of two qubits are therefore mapped to four energy eigenstates of an atom: $|1\rangle$, $|2\rangle$, $|3\rangle$, $|4\rangle$ respectively. For example, the entangled state $|\Phi^+\rangle = 1/\sqrt2 (|00\rangle+|11\rangle)$, is  encoded in a superposition of states of four level atom as $1/\sqrt2 (|1\rangle+|4\rangle)$. With that assignment we have
\begin{equation}
    \tilde W=|1\rangle\langle4| + |4\rangle\langle1|,
\end{equation}
which is consistent with (\ref{Ham4}). \textcolor{black}{To let the interaction depend on either $W_+$ or $W_-$ we displace the original modes as $a_\alpha=D(\alpha)aD^\dagger(\alpha)$. Assuming $\alpha$ to be real, the old modes are $a=a_\alpha+\alpha$ and $a^\dagger=a^\dagger_\alpha+\alpha$. With this substitution the Hamiltonian (\ref{Ham4}) reads (up to a constant) 
\begin{equation}
      H =\sum_{i=1}^4 \varepsilon_i \sigma_{ii} + 2\alpha\gamma \tilde W+ \omega a^\dagger_{\alpha} a_{\alpha}  + (\alpha\omega+ \gamma \tilde W)(a_{\alpha}+ a^\dagger_{\alpha}).
\end{equation}
By letting $\alpha=|\gamma|/\omega$ the interaction term becomes $|\gamma|W_\pm (a_{|\gamma|/\omega}+ a^\dagger_{|\gamma|/\omega})$, 
where $\pm$ is the sign of $\gamma$. Finally, since we require that $W_\pm$ does not evolve in time, we have to set $\varepsilon_1=\varepsilon_4$.}

\section{Conclusions} 

We address the problem of indirect certification of entanglement in a quantum system, by solely analysing the environment with which this system interacts. The method focuses on performing measurements of the environment observables, the behavior of which generically depends on whether the system of interest exhibits entanglement with respect to its intrinsic degrees of freedom. 
Therefore, it is complementary to ancilla-assisted measurement schemes which have recently  been successfully developed \cite{ar1,ar2,ar3,ar4} in the context of assessing thermodynamic properties \cite{aam1} and in the area of thermometry \cite{aam2}. Our model is based on introducing the Hamiltonian of the system and its environment, with the interaction part consisting of the entanglement witness acting on the space of the investigated system. Given that the system of interest is entangled, the interaction with the environment can lead to emergence of salient signatures which are imprinted in the state of the environment. Consequently, appropriate measurements performed on the environment of the system enable for revealing the entanglement without the need for performing measurements on the system, hence the system itself can be certified to exhibit entanglement. While the method has been designed to cope with true environments, it will be interesting to apply this approach to artificially implemented environments~\cite{Jasmin}. Obviously, there is no universal choice of the entanglement witness: for each particular entangled state to be revealed, ideally one would need to find the optimal entanglement witness \cite{Chruscinski2014}, so that entanglement can be detected most efficiently.

\begin{acknowledgments}
We thank J. Dziewior, L. Knips and J. Meinecke for fruitful discussions.
We acknowledge funding by the Foundation for Polish Science (IRAP project, ICTQT, Contract No. 2018/MAB/5, cofinanced by the EU within the Smart Growth Operational Programme). W.K. and W.L. acknowledge partial support by NCN (Poland) grant no. 2016/23/G/ST2/04273.
\end{acknowledgments}

\appendix

\section{Thermal fluctuations for the ideal gas}
\label{sec:thermalidealgas}

In the main text we found the time-dependent average momentum of the center-of-mass for the ideal gas
\begin{equation}
    P(t) = -\alpha W_0 t.
\end{equation}
Here, we aim to scrutinize its fluctuations around the average, checking whether they might hinder the detection of entanglement in the system. To this end, we shall calculate the standard deviation of the variable considered
\begin{equation}
\label{eq:sigmaP}
    \Sigma_P = \sqrt{\langle P^2 \rangle - P(t)^2} \equiv \sqrt{\langle P^2 \rangle - \left(\alpha W_0 t\right)^2},
\end{equation}
which also will depend on time.

We start by writing the time evolution of the operator $P$ in the Heisenberg picture
\begin{equation}
    U^\dagger(t)\left ( \1_S \otimes P \right ) U(t) = e^{i t H}\left ( \1_S \otimes P \right ) e^{-i t H},
\end{equation}
where $U(t)$ encodes total evolution, so that $H = H_S\otimes \1_E + \1_S \otimes H_E + W\otimes h_E$. Considering the environment Hamiltonian to be equal to $H_E = \sum_i p_i^2 / 2m$ and the Baker-Campbell-Hausdorff (BCH) expansion
\begin{eqnarray}
e^{i t H} B e^{-it H} &=& B + it [H,B] + \frac{(it)^2}{2!}[H,[H,B]] \nonumber \\ &+& \frac{(it)^3}{3!}[H,[H,[H,B]]] +\dots,
\end{eqnarray}
one obtains
\begin{equation}
    U^\dagger(t)\left ( \1_S \otimes P \right ) U(t) = \left ( \1_S \otimes P \right) - \alpha t \left ( W \otimes \1_E \right ).
\end{equation}
 This simple result follows because
\begin{equation}
    [H_E,P] = [W,H_S] = 0,
\end{equation}
and
\begin{equation}
    [W\otimes h_E, \1_S \otimes P] = \alpha N W \otimes [R,P] = i \alpha W \otimes \1_E.
\end{equation}

To find the fluctuations we shall compute
\begin{eqnarray}
    U^\dagger(t) \left(\1_S \otimes P^2 \right)U(t) &=& \1_S \otimes P^2 - 2\alpha t \left ( W \otimes P \right ) \nonumber \\ &+& \alpha^2 t^2 (W^2 \otimes \1_E).
\end{eqnarray}
Taking the average, we use the previous result to get
\begin{eqnarray}
    \langle P^2 \rangle &=&\text{Tr}\left\{\left(\1_S \otimes P^2\right)\rho(t)\right\} \nonumber \\ 
    &=&\langle U^\dagger(t)\left ( \1_S \otimes P^2 \right ) U(t) \rangle_{0},
\end{eqnarray}
where the average on the right hand side is taken at the initial point in time $t=0$, in which the total state of the ``system plus environment'' was a product state, and the environment was  in a thermal state given by the Gibbs form ($\beta=1/k_B T$)
\begin{equation}
 \rho_E(0) = \rho_E^\text{th} = \frac{e^{-\beta H_E}}{Z_E}.
 \end{equation}
 We therefore find that 
\begin{equation}
\label{eq:diffP2afterBCH}
        \langle P^2 \rangle = \langle P^2\rangle_{\text{th}} - 2 \alpha  W_0  \langle P \rangle_{\text{th}} t+ \alpha^2  \langle W^2 \rangle t^2.
\end{equation}
 The momentum average in the second term in Eq.~(\ref{eq:diffP2afterBCH}) is equal to zero. This result applies regardless of the volume occupied by the gas and holds for both quantum and classical statistical description of it. 

While the environment has so far been treated as quantum, for the sake of clarity we perform the thermal average in the classical limit. Then, from 1-dimensional Kinetic Theory, one knows that:
\begin{equation}
    \langle P^2 \rangle_{\text{th}} = \frac{m}{\beta N}.
\end{equation}
Thus:
\begin{equation}
    \langle P^2 \rangle = \frac{m}{N\beta} + \alpha^2 \langle W^2 \rangle t^2.
\end{equation}
Inserting this result into Eq.~(\ref{eq:sigmaP}) gives
\begin{equation}
    \Sigma_P = \sqrt{\alpha^2 \Sigma_W^2 t^2 + \frac{m}{N\beta}},
\end{equation}
where $\Sigma_W$ is the standard deviation of the entanglement witness.

\section{A system coupled to a radiation field}
\label{sec:EMmodes}

To increase the readability of this quite technical section, for the convenience of the reader we are going to use a hat to denote operators describing the electromagnetic field. Still, we are neither applying this rule to Hamiltonians, nor to density matrices. In the main text we completely avoid hats from pragmatic reasons.

We consider a system interacting with a bath consisting of several electromagnetic modes. In such a case, the Hamiltonian of the environment is given by ($\hbar=1=c$)
\begin{equation}
    H_E = \sum_{\boldsymbol k,\lambda} \omega_{k} \left(\hat a^\dagger_\lambda(\boldsymbol k) \hat a_\lambda(\boldsymbol k)+\frac{1}{2}\right),
\end{equation}
where $\boldsymbol k$ is the wave vector and $\omega_{k} = |\boldsymbol k| = k$  is the frequency corresponding to the mode $\boldsymbol k$. As usual, $\hat a_\lambda(\boldsymbol k)$ and $\hat a^\dagger_\lambda(\boldsymbol k)$ are the annihilation and creation operators of the mode $\boldsymbol k$ with polarization $\lambda$. 

The above Hamiltonian stems from quantization of the electromagnetic field in a finite volume $V$, which in the quantum description is represented by the operators in the Schr{\"o}dinger picture:
\begin{eqnarray}
   \hat{\boldsymbol E}(\boldsymbol{r}) &=& i \sqrt{\frac{1}{2 V\epsilon_0}} \sum_{\boldsymbol k,\lambda} \sqrt{\omega_{k}} ( \boldsymbol e _\lambda (\boldsymbol k) \hat a_\lambda(\boldsymbol k)e^{i\boldsymbol{k}\cdot\boldsymbol{r}} \nonumber \\ &-&  
 \boldsymbol e ^*_\lambda (\boldsymbol k) \hat a^\dagger_\lambda(\boldsymbol k)e^{-i\boldsymbol{k}\cdot\boldsymbol{r}} ),
\end{eqnarray}
\begin{eqnarray}
    \hat{\boldsymbol B}(\boldsymbol{r}) &=&  i\sqrt{\frac{1}{2 V\epsilon_0}} \sum_{\boldsymbol k,\lambda} \frac{1}{\sqrt{\omega_{k}}}\boldsymbol{k}\times ( \boldsymbol e _\lambda (\boldsymbol k) \hat a_\lambda(\boldsymbol k)e^{i\boldsymbol{k}\cdot\boldsymbol{r}} \nonumber \\ &-&
 \boldsymbol e ^*_\lambda (\boldsymbol k)\hat a^\dagger_\lambda(\boldsymbol k)e^{-i\boldsymbol{k}\cdot\boldsymbol{r}} ).
\end{eqnarray}
Polarization vectors $\boldsymbol e_\lambda (\boldsymbol k)$ together with the wave vector $\boldsymbol k$ satisfy the following relations \cite{Breuer2002}:
\begin{equation}\label{trans}
    \boldsymbol k \cdot \boldsymbol e _\lambda (\boldsymbol k) = 0,
\end{equation}
\begin{equation}
    \boldsymbol e _\lambda (\boldsymbol k) \cdot \boldsymbol e _{\lambda '} (\boldsymbol k) = \delta_{\lambda \lambda'},
\end{equation}
\begin{equation}\label{lambda}
    \sum_{\lambda=1,2} e^i _\lambda (\boldsymbol k) \left[e^j _\lambda (\boldsymbol k)\right]^* = \delta_{ij} - \frac{k_i k_j}{k^2} \quad i,j = 1,2,3.
\end{equation}
The indices $i,j$ denote the components of the corresponding vector in Cartesian coordinates.

We choose the interaction term $h_E$ to be equal to
\begin{equation}
h_E = -\int \dd^3\boldsymbol{r}\boldsymbol D (\boldsymbol{r})\cdot \hat{\boldsymbol E}(\boldsymbol{r}),
\end{equation}
with $\boldsymbol D(\boldsymbol{r})$ being a given real vector representing the dipole moment. We find
\begin{eqnarray}
h_E &=& -i \sqrt{\frac{1}{2 V\epsilon_0}} \sum_{\boldsymbol k,\lambda} \sqrt{\omega_{k}} (\boldsymbol d (\boldsymbol{k})\cdot \boldsymbol e _\lambda (\boldsymbol k) \hat a_\lambda(\boldsymbol k) \nonumber \\ &-& 
\boldsymbol d^* (\boldsymbol{k})\cdot \boldsymbol e ^*_\lambda (\boldsymbol k)\hat a^\dagger_\lambda(\boldsymbol k) ),
\end{eqnarray}
where
\begin{eqnarray}\label{Fourier}
    \boldsymbol{d}(\boldsymbol{k}) &=& \int \dd^3\boldsymbol{r}\boldsymbol D (\boldsymbol{r}) e^{i\boldsymbol{k}\cdot\boldsymbol{r}}, \nonumber \\ \boldsymbol D (\boldsymbol{r})&=&\frac{1}{(2\pi)^3} \int \dd^3\boldsymbol{k} \boldsymbol{d}(\boldsymbol{k})  e^{-i\boldsymbol{k}\cdot\boldsymbol{r}}.
\end{eqnarray}

As explained in the main text, time evolution of averages of environment operators depends on the forces given through commutators with $H_E$ and $h_E$. Therefore, the aim would be to calculate these forces, with the variables $G_j$ given by both $\hat{\boldsymbol E}(\boldsymbol{r})$ and $\hat{\boldsymbol B}(\boldsymbol{r})$. However, to let the presentation be technically simpler (no need to consider field-theoretic commutators of the field operators) we apply this methodology to all creation and annihilation operators. We obtain:
\begin{equation}
    [ H_E, \hat a_\lambda (\boldsymbol k) ] = -\omega_{k} \hat a_\lambda (\boldsymbol k),
\end{equation}
\begin{equation}
    [ H_E, \hat a^\dagger_\lambda (\boldsymbol k) ] = \omega_{k} \hat a^\dagger_\lambda (\boldsymbol k),
\end{equation}
\hfill
\begin{equation}
    [ h_E, \hat a_\lambda (\boldsymbol k) ]  = i \Omega_\lambda(\boldsymbol k),\qquad  [ h_E, \hat a^\dagger_\lambda (\boldsymbol k) ] = i \Omega^*_\lambda(\boldsymbol k),
\end{equation}
with 
\begin{equation}
    \Omega_\lambda(\boldsymbol k) =- \sqrt{\frac{\omega_{k}}{2V\epsilon_0}}\boldsymbol d^*(\boldsymbol{k}) \cdot \boldsymbol e^*_\lambda(\boldsymbol k).
\end{equation}
Putting everything together, the differential equations that dictate the time evolution of the average values of the creation and annihilation operators are:
\begin{equation}
    \frac{\dd}{\dd t} \langle \hat a_\lambda (\boldsymbol k)\rangle = -i\omega_{k} \langle \hat a_\lambda (\boldsymbol k)\rangle - \Omega_\lambda(\boldsymbol k) W_0,
\end{equation}
\begin{equation}
    \frac{\dd}{\dd t} \langle \hat a^\dagger_\lambda (\boldsymbol k)\rangle = i\omega_{k} \langle \hat a^\dagger_\lambda (\boldsymbol k)\rangle - \Omega^*_\lambda(\boldsymbol k) W_0.
\end{equation}
The solutions of these equations are:
\begin{eqnarray}
    \langle \hat a_\lambda (\boldsymbol k)\rangle_t &=& \langle \hat a_\lambda (\boldsymbol k)\rangle_0 e^{-i \omega_{k} t} \nonumber \\ &+& \frac{i}{\omega_{\boldsymbol k}}\Omega_\lambda(\boldsymbol k) W_0 (1-e^{-i \omega_{k} t}),
\end{eqnarray}
\begin{eqnarray}
    \langle \hat a^\dagger_\lambda (\boldsymbol k)\rangle_t &=& \langle \hat a^\dagger_\lambda (\boldsymbol k)\rangle_0 e^{i \omega_{k} t} \nonumber \\ &-& \frac{i}{\omega_{\boldsymbol k}}\Omega^*_\lambda(\boldsymbol k) W_0 (1-e^{i \omega_{k} t}).
\end{eqnarray}

In the next step, we assume that the electromagnetic bath is initially in a thermal state
\begin{equation}
    \rho_E(0) = \frac{e^{-\beta H_E}}{Z_E}.
\end{equation}
Then, the initial average values of the annihilation and creation operators vanish
\begin{equation}
    \langle \hat a_\lambda (\boldsymbol k)\rangle_0 = 0=\langle \hat a^\dagger_\lambda (\boldsymbol k)\rangle_0,
\end{equation}
so, since
\begin{eqnarray}
    \boldsymbol E(\boldsymbol{r},t)  &=& i \sqrt{\frac{1}{2 V\epsilon_0}} \sum_{\boldsymbol k,\lambda} \sqrt{\omega_{k}} ( \boldsymbol e _\lambda (\boldsymbol k)  \langle \hat a_\lambda (\boldsymbol k)\rangle_t e^{i\boldsymbol{k}\cdot\boldsymbol{r}} 
    \nonumber \\ &-& 
 \boldsymbol e ^*_\lambda (\boldsymbol k)  \langle \hat a^\dagger_\lambda (\boldsymbol k)\rangle_t e^{-i\boldsymbol{k}\cdot\boldsymbol{r}} ),
\end{eqnarray}
after a couple of simplifications we get
\begin{eqnarray}
    \boldsymbol E(\boldsymbol{r},t) &=& - W_0\sqrt{\frac{2}{ V\epsilon_0}} \sum_{\boldsymbol k,\lambda} \frac{1}{\sqrt{\omega_{k}}}  \\ &\times& \mathrm{Re}\left[ \boldsymbol e _\lambda (\boldsymbol k) \Omega_\lambda(\boldsymbol k)e^{i\boldsymbol{k}\cdot\boldsymbol{r}} (1-e^{-i \omega_{k} t}) \right],\nonumber
\end{eqnarray}
Inserting the definition of $\Omega_\lambda(\boldsymbol k)$, summing over $\lambda$ according to Eq. (\ref{lambda}) and performing the transformation between the sum over $\boldsymbol k$ and the integral over $\dd^3\boldsymbol k$ 
\begin{equation}
    \sum_{\boldsymbol k} \mapsto \frac{V}{(2\pi)^3}\int \dd^3 \boldsymbol k,
\end{equation}
we obtain
\begin{eqnarray}
    \boldsymbol E(\boldsymbol{r},t) &=&  W_0\frac{1}{(2\pi)^3 \epsilon_0} \int \dd^3 \boldsymbol k \\ &\times& \mathrm{Re}\left[\left(\boldsymbol d(\boldsymbol{k}) - \frac{\boldsymbol{k}\cdot \boldsymbol d(\boldsymbol{k})}{k^2}\boldsymbol{k} \right) e^{-i\boldsymbol{k}\cdot\boldsymbol{r}} (1-e^{i \omega_{k} t}) \right]. \nonumber
\end{eqnarray}

According to Maxwell equations the electric field operator is transversal, which is best captured by the relation (\ref{trans}). For the sake of clarity we now also assume that the dipole moment is transversal as well, so that $\boldsymbol{k}\cdot \boldsymbol d(\boldsymbol{k})=0$. In such a case
\begin{equation}
    \boldsymbol E(\boldsymbol{r},t) =  W_0\frac{1}{(2\pi)^3 \epsilon_0} \int \dd^3 \boldsymbol k \mathrm{Re}\left[\boldsymbol d(\boldsymbol{k})  e^{-i\boldsymbol{k}\cdot\boldsymbol{r}} (1-e^{i \omega_{k} t}) \right].
\end{equation}
Before going further we also observe that the average magnetic field in this situation is equal to
\begin{eqnarray}
    \boldsymbol B(\boldsymbol{r},t) &=&  W_0\frac{1}{(2\pi)^3 \epsilon_0} \int \dd^3 \boldsymbol k \\ &\times& \mathrm{Re}\left[\frac{1}{\omega_k}\boldsymbol{k}\times\boldsymbol d(\boldsymbol{k})  e^{-i\boldsymbol{k}\cdot\boldsymbol{r}} (1-e^{i \omega_{k} t}) \right]. \nonumber
\end{eqnarray}
In fact, these two averages are our macroscopic variables $G_j(t)$.

In the next step, we average the above electromagnetic field over time:
\begin{eqnarray}
  \overline{\boldsymbol E(\boldsymbol{r})} &=&\lim_{T\rightarrow\infty} \frac{1}{T} \int_0^T \dd t \, \boldsymbol E(\boldsymbol{r},t), \nonumber \\ \overline{\boldsymbol B(\boldsymbol{r})} &=&\lim_{T\rightarrow\infty} \frac{1}{T} \int_0^T \dd t \, \boldsymbol B(\boldsymbol{r},t),
\end{eqnarray}
getting:
\begin{equation}
    \overline{\boldsymbol E(\boldsymbol{r})} =  W_0\frac{1}{(2\pi)^3 \epsilon_0} \int \dd^3 \boldsymbol k \mathrm{Re}\left[\boldsymbol d(\boldsymbol{k})  e^{-i\boldsymbol{k}\cdot\boldsymbol{r}} \right],
\end{equation}
\begin{equation}
   \overline{\boldsymbol B(\boldsymbol{r})} =  W_0\frac{1}{(2\pi)^3 \epsilon_0} \int \dd^3 \boldsymbol k \mathrm{Re}\left[\frac{1}{\omega_k}\boldsymbol{k}\times\boldsymbol d(\boldsymbol{k})  e^{-i\boldsymbol{k}\cdot\boldsymbol{r}}  \right].
\end{equation}

Replacing $\boldsymbol{k}\times$ by $i\boldsymbol{\nabla}\times$ in the time-averaged magnetic field, and taking into account that $\boldsymbol D(\boldsymbol{r})$ is real, we find that 
\begin{equation}
   \overline{\boldsymbol B(\boldsymbol{r})} =  0,
\end{equation}
and
\begin{equation}
    \overline{\boldsymbol E(\boldsymbol{r})} =  \frac{W_0}{\epsilon_0} \boldsymbol D(\boldsymbol{r}).
\end{equation}
This is our final result.

\providecommand{\noopsort}[1]{}\providecommand{\singleletter}[1]{#1}%

\end{document}